\documentclass[noshowpacs,amsmath,twocolumn,
aps,prb]
{revtex4-1}

\usepackage{setspace}
\usepackage{amsmath}
\usepackage{graphicx}
\usepackage[nearskip,margin = 0pt]{subfig}
\usepackage{subfig}

\usepackage{verbatim}
\usepackage{amsfonts}
\usepackage{amssymb}
\usepackage{epstopdf} 
\usepackage{xcolor}

\DeclareGraphicsExtensions{.pdf,.eps,.png,.jpg,.mps} 

\begin{document}

\title{Searching for Exoplanets Using a Microresonator Astrocomb }


\author{ Myoung-Gyun Suh$^{1}$, Xu Yi$^{1}$, Yu-Hung Lai$^{1}$, S. Leifer$^{2}$, Ivan S. Grudinin$^{2}$, G. Vasisht$^{2}$, Emily C. Martin$^{4}$, Michael P. Fitzgerald$^{4}$, G. Doppmann$^{5}$, J. Wang$^{6}$, D. Mawet $^{6,2}$, Scott B. Papp$^{3}$, Scott A. Diddams$^{3}$, C. Beichman$^{7,*}$, Kerry Vahala$^{1,*}$\\
$^1$T. J. Watson Laboratory of Applied Physics, California Institute of Technology, Pasadena, California 91125, USA\\
$^2$Jet Propulsion Laboratory, California Institute of Technology, Pasadena, California 91109, USA\\
$^3$National Institute of Standards and Technology, 325 Broadway, Boulder, Colorado 80305, USA\\
$^4$Department of Physics and Astronomy, University of  California Los Angeles, Los Angeles, CA 90095, USA\\
$^5$W.M. Keck Observatory, Kamuela, HI 96743, USA\\
$^6$Department of Astronomy, California Institute of Technology, Pasadena, CA 91125, USA\\
$^7$NASA Exoplanet Science Institute, California Institute of Technology, Pasadena, CA 91125, USA\\
$^*$Corresponding author: Kerry Vahala (email: vahala@caltech.edu) and C. Beichman (email: chas@ipac.caltech.edu)
}

\maketitle
\newcommand{\ts}{\textsuperscript}
\newcommand{\tsb}{\textsubscript}



{\bf Detection of weak radial velocity shifts of host stars induced by orbiting planets is an important technique for discovering and characterizing planets beyond our solar system. Optical frequency combs enable calibration of stellar radial velocity shifts at levels required for detection of Earth analogs. A new chip-based device, the Kerr soliton microcomb, has properties ideal for ubiquitous application outside the lab and even in future space-borne instruments. Moreover, microcomb spectra are ideally suited for astronomical spectrograph calibration and eliminate filtering steps required by conventional mode-locked-laser frequency combs. Here, for the calibration of astronomical spectrographs, we demonstrate an atomic/molecular line-referenced, near-infrared soliton microcomb. Efforts to search for the known exoplanet HD 187123b were conducted at the Keck-II telescope as a first in-the-field demonstration of microcombs. 


}


A fundamental question of the human race, whether life exists on other planets, has brought huge interest in searching for Earth-like extrasolar planets (exoplanets), especially in the `habitable zone'\cite{kasting1993habitable} where the orbital separation is suitable for the presence of liquid water at the planet's surface. Since the discovery of an exoplanet around a solar-type star\cite{campbell1988search,mayor1995jupiter}, thousands of exoplanets have been reported\cite{akeson2017nasa}. In this quest, the radial velocity (RV) method (Figure 1) employs high-precision spectroscopic measurements of periodic Doppler shifts in the stellar spectrum to infer the presence of an orbiting exoplanet \cite{perryman2011exoplanet}. Importantly, the RV technique provides information about the exoplanet mass, which is unavailable with the complementary technique of transit photometry. However, RV detection of an earth-like planet in the habitable zone requires extreme spectral precision of about $3\times10^{-10}$, equivalent to a recoil velocity of the star of only $\sim 10 $ cm s$^{-1}$.

\begin{figure}[t!]
  \begin{centering}
  \includegraphics[width=\linewidth]{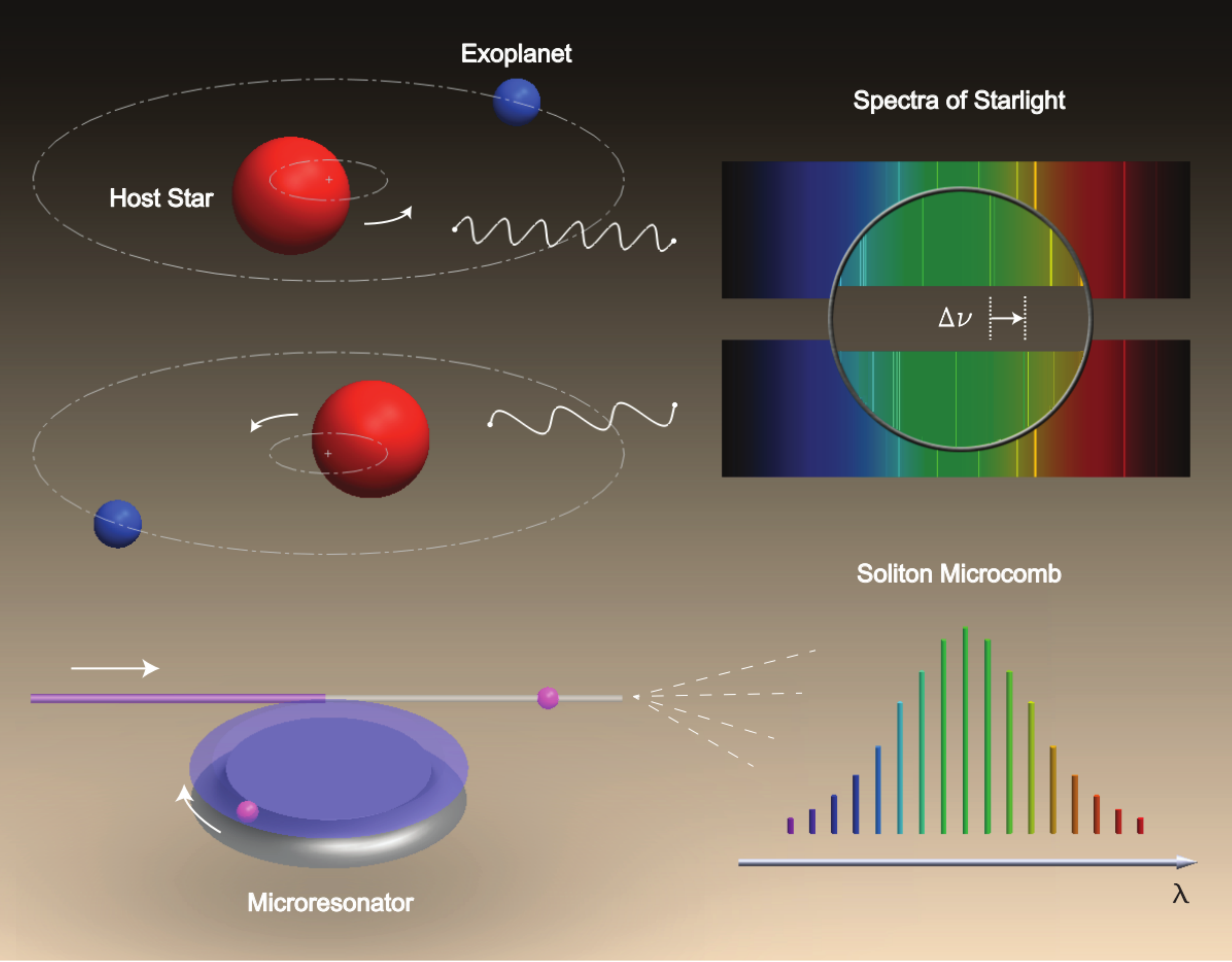}
    \captionsetup{singlelinecheck=no, justification = RaggedRight}
    \caption{{\bf Concept of a microresonator astrocomb.} While the host star (red sphere) and exoplanet (blue sphere) orbit their common center of mass, light waves leaving the star experience a weak Doppler shift. The frequency shift ($\Delta \nu$) of the stellar spectral lines are measured using a spectrograph calibrated using an evenly spaced comb of frequencies. Here, the comb of frequencies is produced by a soliton emission from a microresonator.}
    \label{fig1}
  \end{centering}
\end{figure}

\begin{figure*}[t!]
  \begin{centering}
  \includegraphics[width=\linewidth]{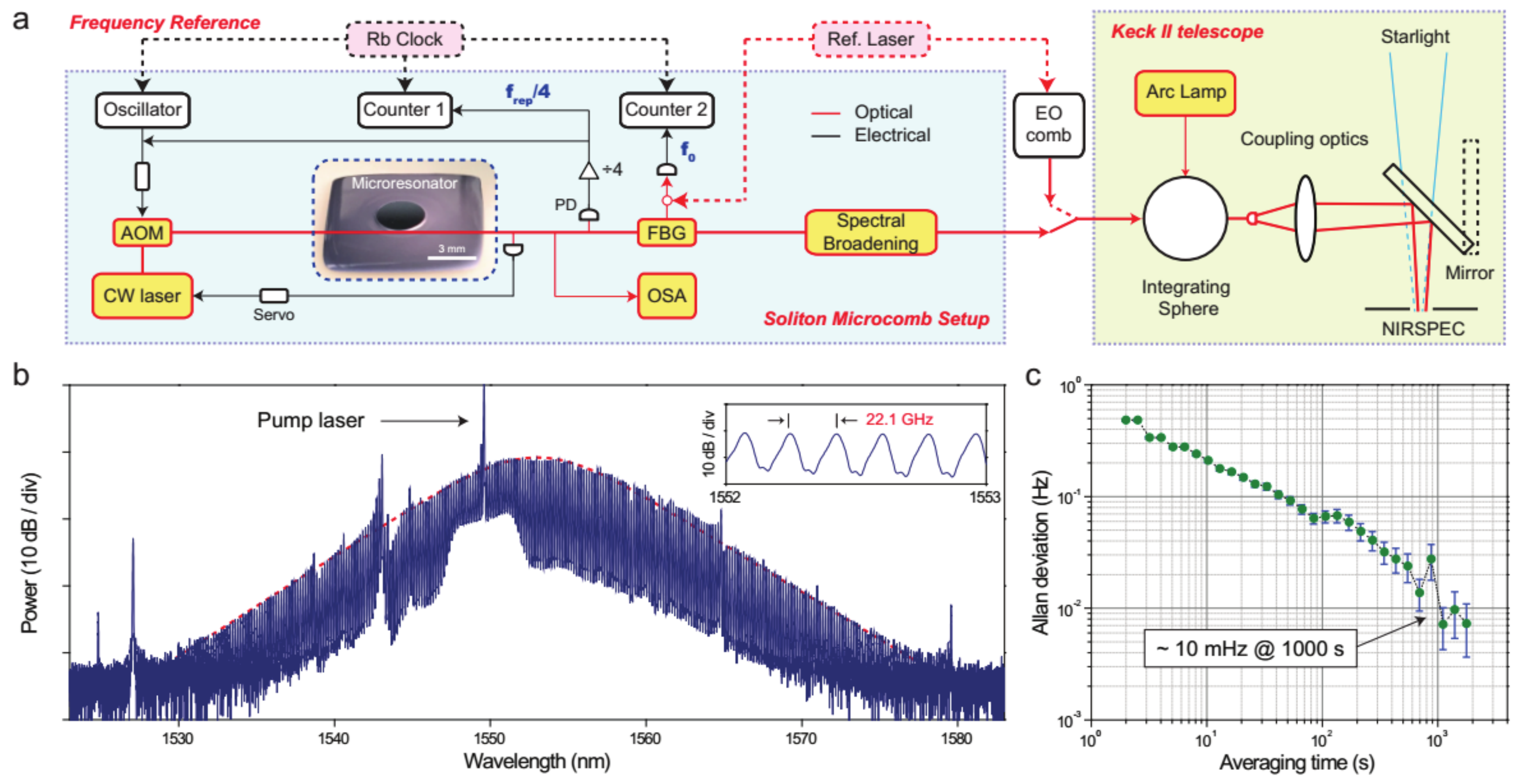}
    \captionsetup{singlelinecheck=no, justification = RaggedRight}
    \caption{{\bf Experimental schematic and atomic/molecular line-referenced soliton microcomb.} (a) Continuous-wave (CW) fiber laser is coupled into a silica microresonator via a tapered fiber coupler\cite{cai2000observation,spillane2003ideality}. An acousto-optic modulator (AOM) controls pump power. The soliton microcomb is long-term stabilized by servo control of the pump laser frequency to hold a fixed soliton average power\cite{yi2016active}. The comb power is also tapped to detect and stabilize the repetition frequency ($f_{rep}$). After dividing by 4, $f_{rep}$ is frequency-locked to an oscillator and monitored using a frequency counter. A rubidium (Rb) clock provides an external frequency reference. The frequency offset ($f_0$) of a soliton comb line is measured relative to a reference laser (stabilized to HCN at 1559.9 nm). This comb line is filtered-out by a fiber Bragg grating (FBG) filter and heterodyned with the reference laser. Finally, the soliton microcomb is spectrally broadened and sent to the integrating sphere of the NIRSPEC instrument on the Keck II telescope for spectrograph calibration. As a cross check, an EO comb (instead of soliton microcomb) is also used. (b) Optical spectrum of the soliton microcomb. The hyperbolic-secant-square fit (red dotted curve) indicates that the soliton pulse width is 145 fs. Inset : Zoom-in of the spectra showing 22.1 GHz line spacing.  (c) Allan deviation of the frequency-locked $f_{rep}/4$. PD : photodetector, OSA : optical spectrum analyzer.}
    \label{fig2}
  \end{centering}
\end{figure*}

\begin{figure*}[t!]
    \begin{centering}
   \includegraphics[width=\linewidth]{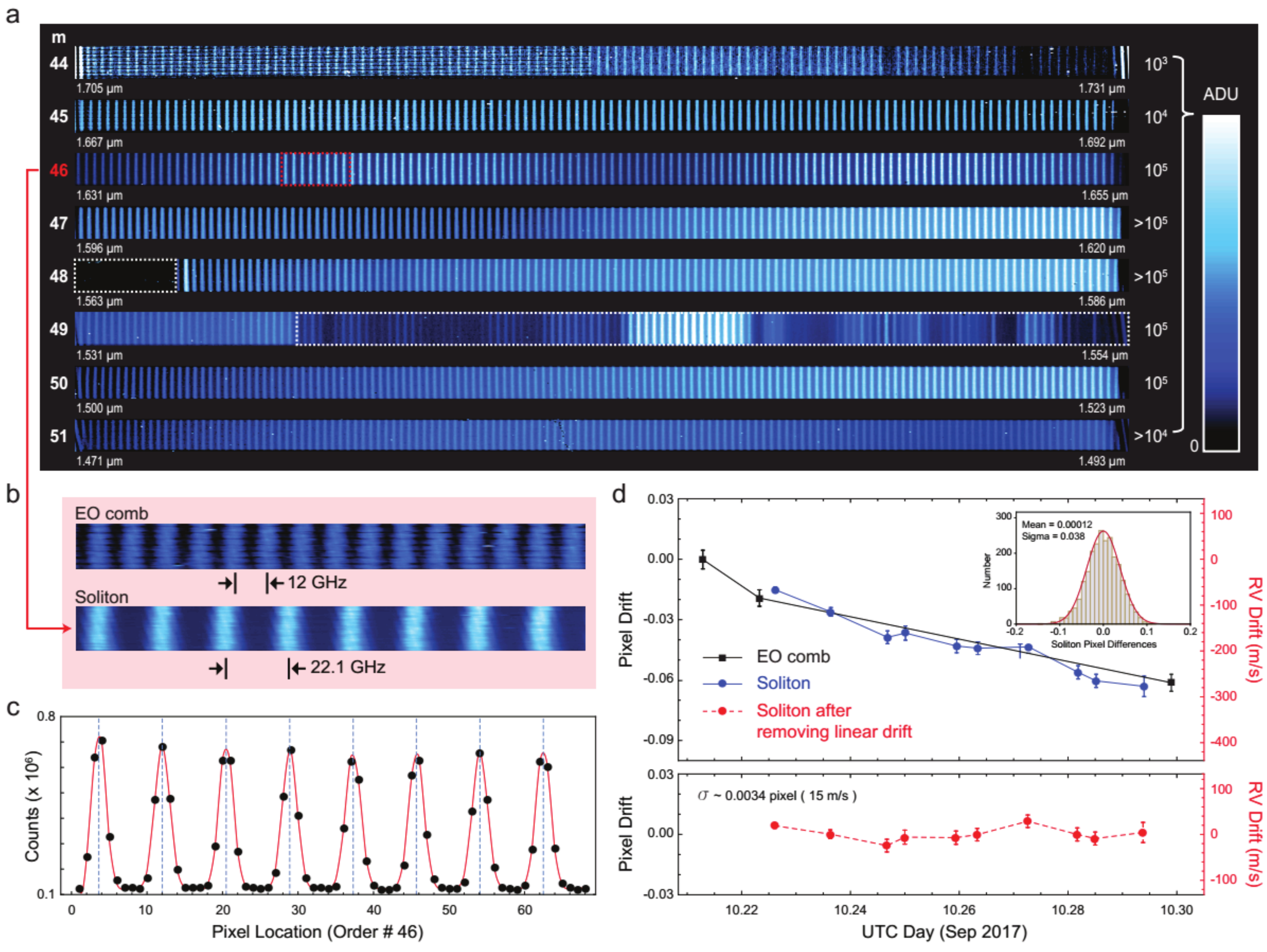}       \captionsetup{singlelinecheck=no, justification = RaggedRight}
        \caption{{\bf Data from testing at Keck II}. (a) Image of soliton comb projected onto the NIRSPEC echelle spectrometer from the echelle orders 44 to 51 with the corresponding wavelength ranges of each order indicated. The white dashed box indicates soliton emission and has been strongly filtered to prevent potential damage of the spectrograph. ADU: Analogue-to-Digital Units. (b) A zoom-in of the Echelle order 46 (red dashed box in panel a) of the EO comb (upper) and soliton (lower) with line spacings indicated. (c) Gaussian profiles with variable amplitude, centroid, $x_j(t)$ and width were fitted (see Methods) to each soliton comb line in the echellogram order (\#46). An illustrative sample of 8 adjacent comb lines is shown. (d) Upper panel shows average centroid drift within Order \#46, $z(t)$, relative to the first frame in the time series with both the soliton (blue) and EO combs (black) during a quiescent portion of the on-sky observing time on 9/10/2017 UTC. Measurements were taken with the EO comb bracketing the soliton comb over a 2.5 hour period. The solid black curve shows the drift of the NIRSPEC wavelength scale in units of a NIRSPEC pixel ($\sim 4$ km  s$^{-1}$) measured with the EO comb. The zero-point of the soliton measurement was adjusted to be approximately the same as the EO comb close to the beginning of the soliton dataset (near UTC=10.22). The lower panel shows the NIRSPEC drift after subtracting a linear trend and gives a residual of 0.0034 pixel which corresponds to approximately 15 m s$^{-1}$ in a single order.   The inset in the upper panel shows that the distribution of centroid differences is well defined by a Gaussian distribution (see the Methods section). As discussed in the text the final wavelength calibration across the entire echellogram would be $\leq5$ m s$^{-1}$.} 
    \end{centering}
\end{figure*}

Paramount, therefore, to RV measurements is a precisely-calibrated astronomical spectrometer\cite{pepe2014instrumentation}. A powerful, new calibration technique is the use of laser frequency combs (LFCs) to provide a spectrally broad `comb' of optical frequencies that are precisely stabilized through the process of self-referencing\cite{Jones2000}. Self-referencing ensures that both the comb's spectral line spacing and the common offset frequency of the spectral lines from the origin are locked to a radio frequency standard resulting in a remarkably accurate `optical ruler'. LFCs have revolutionized spectroscopy, time standards, microwave generation and a wide range of other applications\cite{diddams2010evolving}. For RV detection of exoplanets, LFCs or simply astrocombs have enabled spectrograph calibration that could enable RV detection at the cm s$^{-1}$ level\cite{wilken2012spectrograph}. This is both well below requirements for detection of Earth-like planets in the habitable zone of sun-like stars, and also at a level not achievable by conventional calibration methods using the emission lines of hollow-cathode gas lamps. In many cases, however, telluric atmospheric effects and/or intrinsic stellar noise will impose detection limits above this level. 

The astrocombs used in these earlier experiments\cite{murphy2007high,li2008laser,braje2008astronomical,wilken2012spectrograph,glenday2015operation,mccracken2017decade,steinmetz2008laser} are derived from femtosecond modelocked lasers that feature a comb line frequency spacing that is not resolvable by most astronomical spectrographs \cite{diddams2010evolving}. This has necessitated the addition of special spectral filters designed to coarsen the line spacing \cite{murphy2007high,li2008laser,braje2008astronomical,wilken2012spectrograph,glenday2015operation,mccracken2017decade,steinmetz2008laser}. The added complexity of this filtering step has created interest in frequency comb generation by other means that can intrinsically provide a readily resolvable line spacing. For example, electro-optical (EO) modulation provides an alternative approach for direct generation of $>$10 GHz comb line spacings\cite{murata2000optical,huang2006high}. Line referenced EO-astrocomb devices \cite{yi2016demonstration} and more recently, self-referenced EO-combs\cite{Beha:17,carlson2017ultrafast} have been demonstrated. However, these devices require optical filtering to remove amplified phase noise in the wings of the broadened comb. Another optical source that produces wider comb line spacing is in the form of a tiny microresonator-based comb or microcomb\cite{del2007optical,kippenberg2011microresonator}. Driven by parametric oscillation and four-wave-mixing\cite{kippenberg2004kerr}, millimeter-scale versions of these devices have line spacings that are ideally suited for astronomical calibration\cite{kippenberg2011microresonator}. However, until recently microcombs operated in the so-called modulation instability regime of comb formation\cite{Wabnitz:93} and this severely limited their utility in frequency comb applications. 

The recent demonstration of soliton mode-locking in microresonators represents a major turning point for applications of microcombs\cite{herr2014temporal,yi2015soliton,brasch2016photonic,wang2016intracavity,joshi2016thermally,Grudinin:17}. Also observed in optical fiber\cite{leo2010temporal}, soliton formation ensures highly stable mode locking and reproducible spectral envelopes. For these reasons soliton microcombs are being applied to frequency synthesis\cite{spencer2017integrated}, dual comb spectroscopy\cite{suh2016microresonator,dutt2016chip,pavlov2017soliton}, laser-ranging \cite{trocha2017ultrafast,suh2017microresonator}, and optical communications\cite{marin2017microresonator}. Moreover, their compact (often chip-based) form factor and low operating power can enable implementation of these technologies in remote and mobile environments beyond the research lab, including satellites and other space-borne platforms. In this work, we demonstrate a soliton microcomb as an astronomical spectrograph calibrator. We discuss the experimental setup, laboratory results and efforts to detect a previously known exoplanet (HD 187123b) at the W. M. Keck Observatory.

  The on-site soliton microcomb demonstration was performed at the 10 m Keck II telescope of the W.M. Keck Observatory in order to calibrate the near-infrared spectrometer (NIRSPEC). A secondary goal was to detect the RV signature of the 0.5 M$_{Jup}$ planet orbiting the G3V star HD187123 in a 3.1 day period \cite{feng2015california}. Calibrations and observations were performed during the first half nights of 2017-09-10 and 2017-09-11 (UTC) in the hope of detecting the 70 m s$^{-1} $ semi-amplitude of this planetary signature. As a cross-check, the functionality of the soliton microcomb was compared with a previously-demonstrated line-referenced EO-astrocomb \cite{yi2016demonstration}. The experimental apparatus for both combs was established in the computer room adjacent to the telescope control room. Both combs were active simultaneously and the output of either one could be fed into the integrating sphere at the input to the NIRSPEC calibration subsystem via single-mode fiber. The switch between the two combs could be carried out in the computer room within less than a minute by changing the input to the fiber without disturbing NIRSPEC itself. 

The primary elements of the soliton comb calibration system are detailed in Figure 2a. The LFC light (soliton microcomb or EO comb) is sent to the fiber acquisition unit (light green box) to calibrate the NIRSPEC spectrometer \cite{mclean1998design}. Soliton generation uses a silica microresonator fabricated on a silicon wafer\cite{lee2012chemically}. The resonator featured a 3 mm diameter corresponding to an approximate 22.1 GHz soliton comb line spacing and had an unloaded quality factor of approximately 300 million. Figure 2b shows the optical spectrum of the soliton microcomb. For transport to the observatory, the microresonator was mounted inside a brass package with FC/APC fiber connectors. The package was placed inside an insulated foam box and temperature-controlled using a thermoelectric cooler to stay within an operating range.
The soliton repetition frequency ($f_{rep}$) was locked to a rubidium-stabilized local oscillator by servo control of the pump power using an AOM so as to vary the soliton repetition rate. Allan deviation measurement of the locked and frequency-divided signal shows an instability of 10 mHz at 1000 s averaging time (Figure 2c). The frequency of one of the soliton comb lines is monitored by heterodyne detection with a reference laser, which is locked to a hydrogen cyanide (HCN) absorption line at 1559.9 nm. The resulting offset frequency $f_{0}$ is recorded at every second using a frequency counter stabilized to the Rb clock with a time stamp for calibration of the frequency comb over time. For calibration of the frequency comb over time, $f_0$ was determined over a 20 second averaging time (i.e., acquisition time for a single spectrum) with standard deviation less than 1 MHz. Over this time, the absolute optical frequency of the HCN reference laser has an imprecision of less than 1 MHz \cite{yi2016demonstration}. Because the soliton repetition rate (i.e., comb line spacing) is frequency locked, the offset frequency imprecision was the principal source of instability in the comb calibration, equivalent to about 1 m s$^{-1}$ of RV imprecision.  Finally, the soliton microcomb is spectrally broadened using highly nonlinear optical fiber.

Figure 3a shows the echellogram of the soliton microcomb measured by NIRSPEC (8 Echelle orders ranging from 1471 nm to 1731 nm, which represents almost the entire astronomical \textit{H}-band). The raw echellograms were rectified spatially and spectrally. Zoomed-in images of a single order from both the soliton and EO comb data (Figure 3b) show that individual comb lines are resolved at the NIRSPEC resolution of R$\sim$25,000 and spaced approximately 4 pixels apart (0.1 nm) for the EO comb and 8 pixels (0.2 nm) for the soliton comb.



The soliton and EO comb time series data consisted of hundreds of frames taken every 20 seconds over the course of many hours. As described in the Methods section, the reduced echellograms for the two combs were analyzed in a similar manner by fitting a Gaussian to each line (Figure 3c) to determine its pixel location. For this analysis we chose Order \#46 which spans 1.631 to 1.655 $\mu$m. The centroids of each comb line, $x_j$, were determined across a 2.5 hr interval when the telescope and instrument were in a quiescent state  and investigated for small drifts.  The average drift for the entire order, $z(t)$, consisting of $N=$ 122 (225) lines  in the soliton (EO comb) dataset, was computed with respect to the first frame in the time series and examined as a function of time to reveal drifts within NIRSPEC. 


In the absence of external disturbances such as telescope-induced vibrations, the drift measured continuously in $\sim$ 5 to 10 minute intervals over 2 hours was extremely regular and could be removed by a simple first-order fit. Subtracting the linear drift from the soliton data in the upper panel of Figure 3d results in the red dashed line (lower panel). The soliton comb data reduced the wavelength drift over the two hour interval from 0.027$\pm$ 0.002  pixel  hr$^{-1}$ (120$\pm$10  m s$^{-1} $ hr$^{-1}$)  to zero $\pm$ 0.002 pixel  hr$^{-1}$ ($\pm$10 m s$^{-1} $ hr$^{-1}$). The 1 $\sigma$ residual  around the linear fit in Figure 3d is 0.0034 pixels or 15 m s$^{-1}$.  Other soliton-only datasets taken during these two days showed  residuals as low as 0.0021 pixels  after removal of a linear fit,  or 9 m s$^{-1}$. These values represent  the difference between two frames so that the wavelength precision in a single frame is $\sqrt{2}$ smaller or 10.6 m s$^{-1}$ and 6.5  m s$^{-1}$. 

The inset in Figure 3d (see the Methods section) demonstrates that the distribution of the differences between comb-line centroids from one time step to the next is well represented by a Gaussian distribution, i.e. the final precision is determined by the centroiding uncertainty and the number of comb lines. The wavelength precision obtained above  is  based only on order \#46,  but there are four other orders in the echellogram with comparable amplitude and number of comb lines (Figure 3a). Adding in these other  lines would improve the wavelength solution  by $\sim\times$2, or roughly 3-5 m s$^{-1}$.  Thus the ability to calibrate NIRSPEC at the few m s$^{-1}$  level has been demonstrated using the soliton microcomb near-infrared technology. {\it We emphasize that this wavelength precision  is inherent to NIRSPEC's resolution and stability and it is only the large number of LFC comb lines and their inherent high precision that have revealed the performance of NIRSPEC at this level.}

  Observations of the target star, HD 187123, and a reference star were each bracketed with soliton comb measurements. The analysis of the stellar spectra  and of telluric  absorption lines within those spectra revealed instrumental variations at the 100 m s$^{-1}$ level (0.025-0.05 pixel) which we attribute to wavelength shifts  within  NIRSPEC  which could not be corrected without simultaneous LFC-stellar data. These were not possible in the present configuration. Some of these shifts were clearly associated with telescope motions between different positions on the sky. Motions of the stellar image within the NIRSPEC entrance slit and/or changes in illumination between the integrating sphere (LFC) and the slit (starlight) can also result in shifts of this order. While planet detection could not be achieved, we were able to measure the two combs sequentially and either one with respect to the arc lamps used for the absolute wavelength calibration of NIRSPEC. 
  
  
  
A funded upgrade presently underway will enhance NIRSPEC's  internal thermal and mechanical stability and future upgrades would enable simultaneous observation of an LFC and a stellar image  stabilized via a single mode fiber using Adaptive Optics. Finally, a new generation of spectrographs  is in development by numerous groups to take advantage of diffraction-limited Adaptive Optics imaging to  enable $R>100,000$ spectral resolution and enhanced image stability using single mode fibers. These new instruments will  take full advantage of the wavelength precision available with a new generation of microresonator astrocombs. 


In summary, we report in-situ astronomical spectrograph calibrations with a soliton microcomb. This enables internal instrumental measurement precision at the few m s$^{-1}$ level when calibrating the NIRSPEC astronomical spectrograph at the W. M. Keck observatory (Keck II). The Kerr soliton microcomb we employ already possesses the desirable qualities of $\sim20$ GHz mode spacing, low noise operation and short pulse generation. And rapidly progressing research in this field has resulted in microcomb spectral broadening and self-referencing with integrated photonics\cite{lamb2017optical}, as well as direct generation at shorter wavelengths\cite{lee2017towards}.  These advances will greatly enhance the microcomb stability and bandwidth, which increases the range of detectable astronomical objects. The current prototype system occupies approximately 1.3 m in a standard instrument rack, but significant effort towards system-level integration\cite{spencer2017integrated}, could  ultimately provide a microcomb system in a chip-integrated package with a footprint measured in centimeters. Such dramatic reduction in size is accompanied by reduced weight and power consumption, which would be an enabling factor for ubiquitous frequency comb precision RV calibrations, and other metrology applications in mobile and space-borne instrumentation.\\

{

Note: The authors would like to draw the readers' attention to another microresonator astrocomb demonstration \cite{obrzud2017astrocomb}, which was reported while preparing this manuscript.\\

\noindent\textbf{Methods}

\medskip

\begin{footnotesize}

We calculate the {\it relative} drift in the NIRSPEC wavelength solution, $z(t)$,  at  time  $t$ by taking the average difference in centroid positions of each comb line ($j$=1 to N) in Order \#46 at time $t$, $x_j(t)$, (Figure 3c) relative to the first frame in the time series, $x_j(t=0)$ as defined in eqn (1).

$$z(t)=\frac{1}{N} \sum_{j=1}^{j=N} (x_j(t)-x_j(t=0)). \, \,    \, \, \, \, \, \, \,(1)  $$

\noindent $z(t)$ with its associated uncertainty, $\sigma(t)/\sqrt{N}$, is shown in Figure 3d as measured by both the EO comb (black line) and the soliton microcomb (blue line). The inset in Figure 3d demonstrates that  the distribution of differences in individual soliton comb-line positions after the subtraction of the mean shift, $(x_j(t)-x_j(t=0))-z(t)$ is well characterized by a Gaussian distribution. The precision in determining the frame-to-frame shift is dominated by the centroiding uncertainty (0.038 pixel in the differences, or 0.038/$\sqrt{2}=0.027$ pixel in a single frame) and the total number of comb lines considered.

\end{footnotesize}


\vspace{3 mm}

\medskip

\noindent \textbf{Acknowledgments}

We wish to thank Josh Schlieder for agreeing to share time between his night (2017-09-11) and our own (2017-09-10)  to allow a longer time baseline for the  observations of HD187123. Prof. Andrew Howard generously observed HD187123 with  the Keck I HIRES instrument at visible wavelengths to determine its RV signature in the weeks immediately before our NIR observations. 

We gratefully acknowledge the support of the entire Keck summit team in making these tests possible.
The authors wish to recognize and acknowledge the very significant cultural role and reverence that the summit of Mauna Kea has always had within the indigenous Hawaiian community. We are most fortunate to have the opportunity to conduct observations from this mountain. The data presented herein were obtained at the W.M. Keck Observatory, which is operated as a scientific partnership among the California Institute of Technology, the University of California and the National Aeronautics and Space Administration. The Observatory was made possible by the generous financial support of the W.M. Keck Foundation. This paper made use of data available in the NASA Exoplanet Archive and the Keck Observatory Archive. We thank  Fred Hadaegh for his  support and encouragement which made  this experiment possible. SD and SP acknowledge support from NIST. We thank David Carlson and Henry Timmers for preparing the highly nonlinear optical fiber. KV, MGS, XY and YHL thank the Kavli Nanoscience Institute and the National Aeronautics and Space Administration for support under KJV.JPLNASA-1-JPL.1459106.

This research was carried out at the Jet Propulsion Laboratory and the California Institute of Technology under a contract with the National Aeronautics and Space Administration and funded through the JPL Research and Technology Development. Copyright 2017 California Institute of Technology. All rights reserved. 

\vspace{1 mm}

\noindent \textbf{Author Information} Correspondence and requests for materials should be addressed to KJV (vahala@caltech.edu ).

\bibliography{main}

\end{document}